\documentstyle[12pt]{article}

\newcommand{\spone}{0.9}  

\newcommand{\sptwo}{1.4}
\newcommand{\spthree}{2.4}

\newcommand{\singlespace}{\edef\baselinestretch{\spone}\Large\normalsize}

\newcommand{\doublespace}{\edef\baselinestretch{\sptwo}\Large\normalsize}
\newcommand{\threespace}{\edef\baselinestretch{\spthree}\Large\normalsize}

\singlespace
\threespace
\doublespace

\begin{document}

\begin{center}
{\bf ANALYTIC DEFINITION OF CURVES AND SURFACES\\
BY PARABOLIC BLENDING}\\
{\bf by\\
A.W. Overhauser\\
Mathematical and Theoretical Sciences Department\\
Scientific Laboratory, Ford Motor Company, Dearborn, Michigan\\
Technical Report No: SL 68-40, May 8, 1968}
\end{center}

\begin{center}
{\bf SUMMARY}
\end{center}

A procedure for interpolating between specified points of a
curve or surface is described.  The method guarantees slope
continuity at all junctions.  A surface panel divided into
$p \times q$ contiguous patches is completely specified by
the coordinates of $(p+1)\times (q+1)$ points.  Each individual
patch, however, depends parametrically on the coordinates of
16 points, allowing shape flexibility and global conformity.
\vspace{100pt}

\begin{flushleft}
APPROVED:\\
$~$\\
$~$\\
$~$\\
J.E. Goldman, Director\\
Scientific Laboratory
\end{flushleft}

\pagebreak

\begin{center}
{\bf I.  Interpolation Between points on a Space Curve}
\end{center}

We assume that a space curve is approximately defined by a
sequence of points $\{\vec{A}, \vec{B}, \vec{C},...\}$, each
of which are vectors, i.e., $\vec{A} \cong A_x, A_y, A_z$
in a Cartesian reference frame.  We propose an interpolation
scheme that defines the curve between each pair of adjacent
points so that the curve is as smooth as possible.  Consider
four adjacent points:  D,E,F,G.
\vspace{1in}
\begin{center}
{\bf Figure 1}
\end{center}
\vspace{1in}

\noindent
The distance t along the chord between E and F defines a scalar
variable that we use to parametrize a curve $\vec{c}(t)$
between E and F.  $\vec{c}(t)$ will be a blend of two parabolas.

The three points $\vec{D},\vec{E},\vec{F}$, define a parabola
$\vec{p}(r)$ as follows.  Let r be the distance along the chord
between $\vec{D}$ and $\vec{F}$.  Let u be the distance
perpendicular to r in the plane defined by $\vec{D},\vec{E},\vec{F}$.
The parabola,
$$  u = \alpha r(d-r),
  \eqno{(1)}
$$
has an axis perpendicular to the line along r.  $d =$ distance
between $\vec{D}$ and $\vec{F}$.  We choose $\alpha$ so that this
parabola passes through $\vec{E}$.  Then $\vec{p}(r)$ can be
taken to be the parabola, Eq.(1).

Similarly, the three points $\vec{E},\vec{F},\vec{G}$, define a
parabola $\vec{q}(s)$ as follows.  Let s be the distance along
the chord between $\vec{E}$ and $\vec{G}$.  Let v be the
distance perpendicular to s in the plane defined by
$\vec{E},\vec{F},\vec{G}$.  The parabola,
$$  v = \beta s(e-s),
   \eqno{(2)}
$$
has an axis perpendicular to the line along s.  $e \equiv$ distance
between $\vec{E}$ and $\vec{G}$.  We choose $\beta$ so that this
parabola passes through $\vec{F}$.  Then $\vec{q}(s)$ can be taken to be
the parabola, Eq.(2).  If $\vec{E},\vec{F}$, and $\vec{G}$ happen
to be collinear, Eq.(2) will be $v \equiv 0$, a straight line.

We now define $c(t)$ as a blend of $p(r)$ and $q(s)$ as follows:
$$  \vec{c}(t) = [1 -(t/t_o)]\vec{p}(r) + (t/t_o)\vec{q})(s).
   \eqno{(3)}
$$
$t_o \equiv$ distance between $\vec{E}$ and $\vec{F}$.
Consequently, the two blending functions [the coefficients
of $\vec{p}(r)$ and $\vec{q}(s)]$ vary linearly between
0 and 1.  Eq.(3) is not completely specified until we define a
relation between t and r and between t and s.  This can be done
only by dropping perpendiculars to the lines along r and s
respectively; so $r = r(t)$ and $s = s(t)$ are defined by this
geometric operation.

In like manner we construct a curve $\vec{c}_i(t_i)$ between
each adjacent pair of points.  It is easy to prove that where
two $\vec{c}_i(t_i)$ connect, their slopes are equal; so the
entire curve will be continuous and smooth.  To see this,
consider Eq.(3) rewritten as follows.
$$  \vec{c}(t) = \vec{p}(t) + (t/t_o)[\vec{q}(t)-p(t)].
   \eqno{(4)}
$$
The slope at point $\vec{E}$ is
$$ (d\vec{c}/dt)_E = (d\vec{p}/dt)_E + (t/t_o)_E
     [d\vec{q}/dt-d\vec{p}/dt]_E + (1/t_o)(\vec{q}-\vec{p})_E.
   \eqno{(5)}
$$
The second term on the right hand side is zero because $t=0$ at E;
and the third term on the right hand side is zero because
$\vec{q} = \vec{p}$ at E.  Consequently,
$$  (d\vec{c}/dt)_E = (d\vec{p}/dt)_E.
   \eqno{(6)}
$$
The slope of $\vec{c}(t)$ at E equals the slope of the
parabola $\vec{p}(r)$, which passes through $\vec{D},\vec{E}$,
and $\vec{F}$.  This same parabola determines the slope of
$\vec{c}$ at E for the curve between $\vec{D}$ and $\vec{E}$
by an identical argument.  Consequently slope continuity is
assured.

Blending of two parabolas is possible only if the interval is
an interior one.  If the curve starts at point $\vec{A}$,
\vspace{0.5in}
\begin{center}
{\bf Figure 2}
\end{center}
\vspace{1in}

\noindent
then interpolation between $\vec{A}$ and $\vec{B}$ should be by
the single parabola, defined as above, through $\vec{A},\vec{B}$,
and $\vec{C}$.  Accordingly, in order to specify carefully
the shape of the curve near its endpoint, the distance between
$\vec{A}$ and $\vec{B}$ should be smaller than that for
interior intervals.  There is no requirement that the points
be equally, or nearly equally spaced.  Obviously point
density should be higher where the curvature is higher.

If an interpolated segment of the curve does not behave
according to some a priori intent, an extra point can be
inserted to provide the required constraint.  Aside from
endpoints discussed above, it should not be necessary to
specify for special treatment interior points at inflections
or cusps, provided the latter are not intended to be
infinitely sharp.  Precision is reproducing such features
can be assured merely by specifying extra points sufficiently
near to these characteristic features.

The curve $\vec{c}(t)$, Eq.(3) or (4), is of cubic order
in the Cartesian coordinates.  Consequently it can adequately
represent an inflection that should occur within an interval.
It should be apparent that the cubic function appropriate
to the interval EF does not pass through points D and G.
(This is an important difference in comparison to simple fit
by cubic polynomials.)  The manner of construction (as a
blend of two parabolas) guarantees that spurious wiggles
will not be introduced, as frequently happens when simple
cubes are forced to pass through four points of a curve.

\begin{center}
{\bf II. Interpolation Between Points on a Space Surface}
\end{center}

We assume that a space surface is defined by a net of points
$\vec{A}_{ij}$, such as intersections of lines in the
figure below.
\vspace{2in}
\begin{center}
{\bf Figure 3}
\end{center}
\vspace{2in}

\noindent
Connecting lines between adjacent points of the net can be
constructed by the blending algorithm described in Sec. I.
Clearly a good surface interpolation for the patch EFGH
must take into account the global shape implied by the
adjacent points, $L, M, N, ... V, W$.
Consider the patch EFGH.  A set of coordinates to which
points on the patch are related can be defined, say
$x$ and $y$.  $x =$ constant (between 0 and 1) defines a
space curve through the four points a,b,c,d which,
in the notation of Sec. I, are points $t = xt_o$ of
curves MN, EF, GH, VU respectively.  The line $\ell$,
between b and c, is then defined by the blending algorithm.
Similarly, a line m is defined by $y =$ constant
(between 0 and 1). Surface point $\vec{z}(x,y)$ can be taken to be
$$  \vec{z}(x,y) = \vec{\ell}(y),
   \eqno{(7)}
$$
$$  \vec{z}(x,y) = \vec{m}(x).
    \eqno{(8)}
$$
These two surfaces will differ slightly since $\ell$ and $m$ will
not in general intersect.  They do coincide, however, on all
network lines.  If computational labor is not a factor, one
may take the average of (7) and (8):
$$  \vec{z}(x,y) = \frac{1}{2}\vec{\ell}(y) +
    \frac{1}{2}\vec{m}(x).
   \eqno{(9)}
$$
An advantage of these surface interpolation schemes is that
all surface slopes will be continuous at boundaries between
patches.  The surface shape of each patch will depend, as it
should, on the behavior of the surface surrounding it.

Parabolic blending along $\ell$, m lines is possible
only for interior patches.  Interpolation for patches at an
edge will employ a single parabola for $\ell$ or m, like the
end interval of Sec. I.  Corner patches will necessitate single
parabolas for both $\ell$ and m lines.  Accordingly good edge
definition requires a narrow spacing for the network lines adjacent
to edges of the surface, as shown in the figure.

An entire surface panel consisting of pq patches is specified by
the coordinates of $(p+1)(q+1)$ points, slightly more than
one point per patch on the average.  The shape of a single
(interior) patch, however, depends on 16 points.  This signifies
both flexibility and global relationship.

The optimum arrangement of network points will naturally depend
on the shape of the panel that is to be represented.  If the panel
has a ridge, for example, the network should be arranged so that
a network line passes along the crest.  Closely spaced network
lines (approximately parallel to the crest) should also be included
to provide adequate definition.

\begin{center}
{\bf III. Analysis}
\end{center}

In order to carry out the procedures described in the preceding
sections it is necessary to have explicit formulas for the
parabolic functions and coordinates.  We need to this only for
one case.
\vspace{0.5in}
\begin{center}
{\bf Figure 4}
\end{center}
\vspace{1in}

\noindent
$\vec{J}$ is the point along DF obtained by dropping a perpendicular
from $\vec{E}$.  If $\vec{J}$ is $\vec{D} + x(\vec{F}-\vec{D})$,
then:
$$  \{ \vec{E} - [\vec{D}+x(\vec{F}-\vec{D})]\} \cdot
       (\vec{F}-\vec{D}) = 0.
  \eqno{(10)}
$$
It follows that
$$  x = (\vec{E}-\vec{D}) \cdot (\vec{F}-\vec{D})/d^2
    \eqno{(11)}
$$
where $d \equiv |\vec{F} - \vec{D}|$.  The equation of a point $\vec{p}$
on the parabola $\vec{p}(r)$ is
$$ \vec{p}(r) = \vec{D} + (r/d)(\vec{F}-\vec{D}) +
    \alpha r (d-r) (\vec{E}-\vec{J}).
  \eqno{(12)}
$$
The coefficient $\alpha$ is determined by requiring 
$\vec{p}(xd)  = \vec{E}$.  The last term of (12) must satisfy,
$$  \alpha xd (d-xd)(\vec{E}-\vec{J}) = \vec{E}-\vec{J}.
   \eqno{(13)}
$$
So,
$$  \alpha = 1/[d^2x(1-x)].
   \eqno{(14)}
$$
The equation of the parabola is now completely specified:
Eq.(12) together with values for $\alpha$ and x, Eqs.(14) and (11).

The only problem that remains is to find the relation between
t and r.  From the figure it follows that
$$  r = xd + t\cos\theta .
  \eqno{(15)}
$$
We have,
$$  \cos\theta = (\vec{F}-\vec{E})\cdot (\vec{F}-\vec{D})/dt_o
   \eqno{(16)}
$$
where $t_o = |\vec{F}-\vec{E}|$.  Consequently, Eq.(15) is
specified completely, so Eq.(12) can be written as a function
of t.

In the alternative event that t is along the chord from
$\vec{D}$ to $\vec{E}$, so that
$t_o = |\vec{E}-\vec{D}|$, we have instead of (15),
$$  r = t (\vec{E}-\vec{D})\cdot (\vec{F}-\vec{D})/dt_o .
  \eqno{(17)}
$$

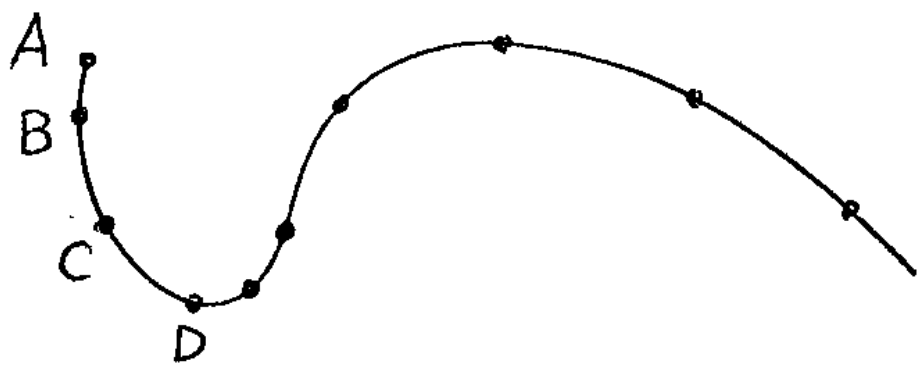
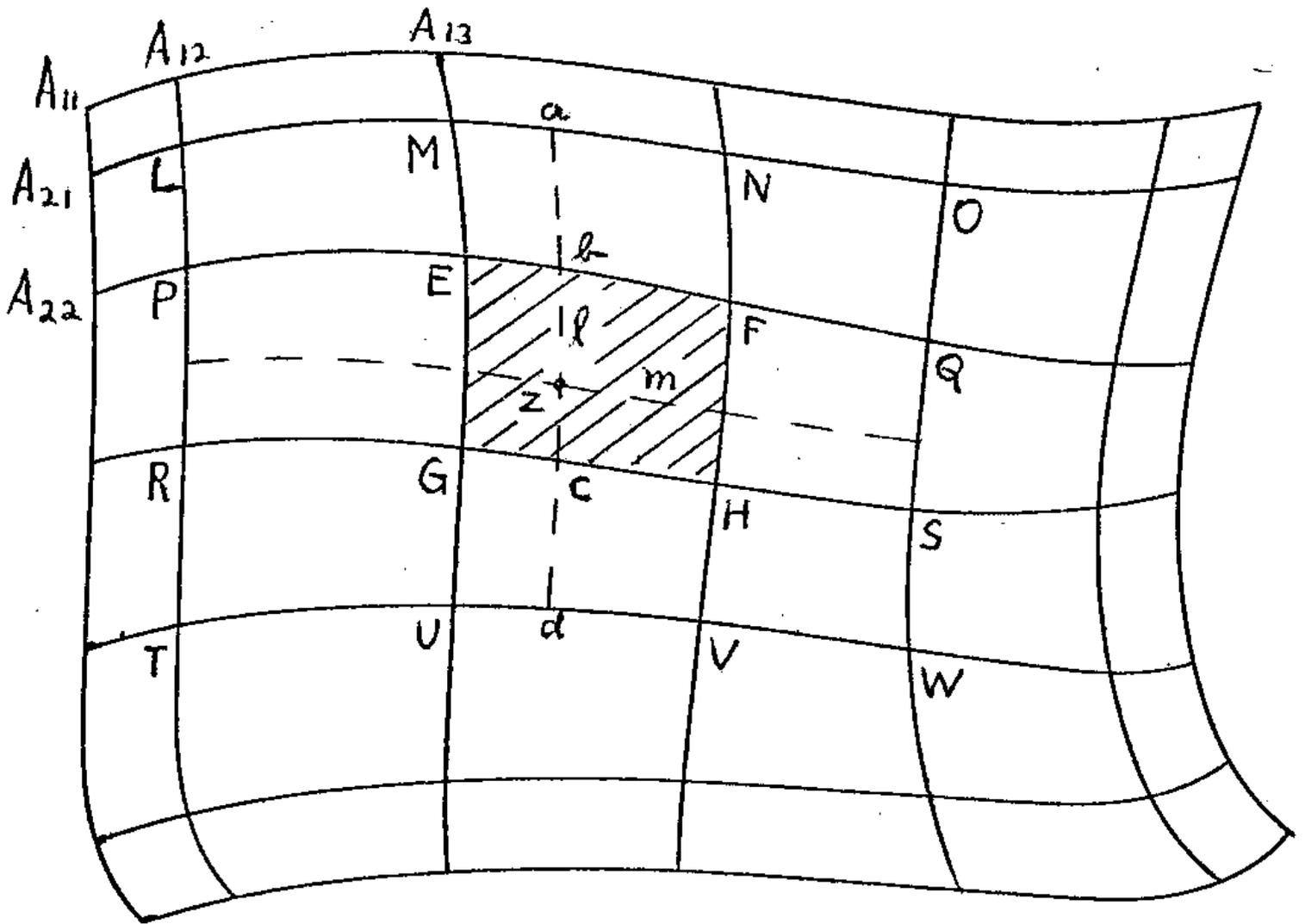
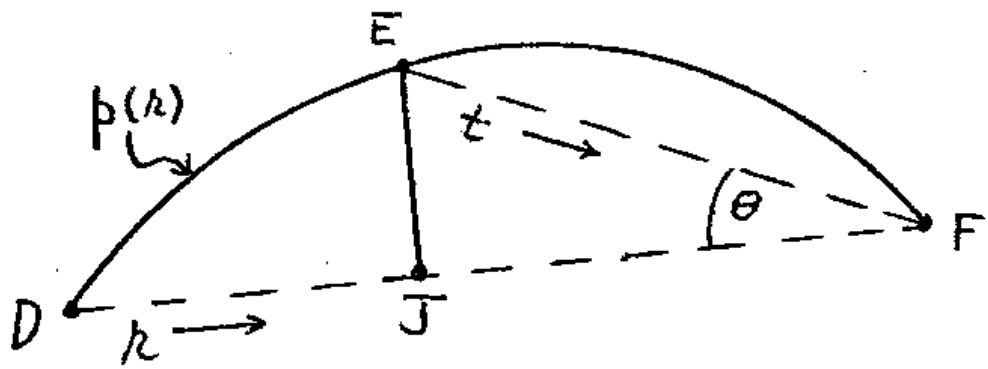
\end{document}